\documentclass[aps,prb,preprint,showpacs,groupaddress,floatfix]{revtex4-1}
\usepackage{graphicx,graphics}
\usepackage{dcolumn}
\usepackage{amsmath,amssymb,amsfonts}
\usepackage{latexsym,verbatim}
\usepackage{bm}
\usepackage{color}
\usepackage{ulem}
\usepackage[breaklinks=true,colorlinks,citecolor=blue,linkcolor=blue,urlcolor=blue]{hyperref}
\def\up{\uparrow}
\def\down{\downarrow}
\def\bef{\begin{framed}}
\def\eef{\end{framed}}
\def\be{\begin{equation}}
\def\ee{\end{equation}}
\def\ber{\begin{eqnarray}}
\def\eer{\end{eqnarray}}

\def\sigmabold{\mbox{\boldmath $\sigma$}}

\def\zv{{\bf \hat z}}
\def\pv{{\bf p}}

\def\xv{{\bf x}}
\def\yv{{\bf y}}

\def\kv{{\bf k}}

\def\Av{{\bf A}}
\def\Bv{{\bf B}}

\def\nn{\nonumber}

\begin{document}
\title{Theory of the nonlinear Rashba-Edelstein effect}
\author{Giovanni Vignale}
\email{vignaleg@missouri.edu}
\affiliation{Department of Physics and Astronomy, University of Missouri, Columbia, Missouri 65211, USA}\affiliation{Donostia International Physics Center (DIPC), Manuel de Lardizabal 4, E-20018 San 
Sebasti\'an, Spain}	
\author{I. V. Tokatly}
\email{ilya.tokatly@ehu.es}
\affiliation{IKERBASQUE, Basque Foundation for Science, E-48011, Bilbao, Spain}
\affiliation{ETSF Scientific Development Centre, 
Departamento de F\'isica de Materiales, Universidad del Pa\'is Vasco, Av. Tolosa 72, E-20018 San 
Sebasti\'an, Spain}
\begin{abstract}
It is well known that a current driven through a two-dimensional electron gas with Rashba spin-orbit coupling induces a spin polarization in the perpendicular direction (Edelstein effect).  This phenomenon has been extensively studied in the linear response regime, i.e., when the average drift velocity of the electrons is a small fraction of the Fermi velocity.   Here we investigate the phenomenon in the nonlinear regime, meaning that the average drift velocity  is comparable to, or exceeds the Fermi velocity.  This regime is realized when the electric field is very large, or when  electron-impurity scattering is very weak.  The quantum kinetic equation for the density matrix of noninteracting electrons is exactly and analytically solvable, reducing to a problem of spin dynamics for ``unpaired" electrons near the Fermi surface.  The crucial parameter is $\gamma=eEL_s/E_F$, where $E$ is the electric field,  $e$ is the absolute value of the electron charge, $E_F$ is the Fermi energy,  and $L_s = \hbar/(
2m\alpha)$ is the spin-precession length in the Rashba spin-orbit field with coupling strength $\alpha$.  If $\gamma\ll1$ the evolution of the spin is adiabatic, resulting in a spin polarization that grows monotonically in time and eventually saturates at the maximum value $n(\alpha/v_F)$, where $n$ is the electron density and $v_F$ is the Fermi velocity.  If $\gamma \gg 1$ the evolution of the spin becomes strongly non-adiabatic and the spin polarization is progressively reduced, and eventually suppressed for $\gamma\to \infty$.  We also predict an inverse nonlinear Edelstein effect, in which an electric current is driven by a  magnetic field that grows linearly in time.  The ``conductivities" for the direct and the inverse effect satisfy generalized Onsager reciprocity relations, which reduce to the standard ones in the linear response regime.      
\end{abstract}
%\pacs{73.20.Mf,71.45.Gm,78.67.Wj}
%
\maketitle
\section{Introduction}
The generation of spin polarization by an electric current and, conversely, of an electric current by a non-equilibrium spin polarization~\cite{Ivchenko78,Levitov85,Aronov89,Edelstein90,KatoEE04,Silov04,Sih05,Silsbee04,Sanchez13,Ganichev02,Sih06,Norman14,Shen14,Shen2-14} are topics of great interest in spintronics~\cite{rmp_76_323,Fabian07,Awschalom07np,MWu09}.    Both effects have a common origin in the spin-orbit interaction and, in the linear response regime, are connected by an Onsager reciprocity relation.  
Experimentally, current-induced spin polarization has been observed in numerous experiments on doped semiconductors~\cite{KatoEE04,Sih05,Sih06,Norman14}.
The inverse effect, known as spin galvanic effect, has been demonstrated in semiconductors~\cite{Ganichev02} and very recently in metallic structures~\cite{Sanchez13}.

On the theoretical side the spin-polarization effect was theoretically predicted in Refs.~\cite{Aronov89,Edelstein90} in the context of the two-dimensional electron gas with Rashba spin-orbit coupling (Rashba 2DEG).  For this reason the effect is widely known in the literature as Rashba-Edelstein effect.  Its inverse was studied theoretically in Ref.~\cite{Levitov85} and, more recently, in Ref.~\cite{Shen14}  (for a more complete discussion see Ref.~\cite{Shen2-14}.  The Edelstein effect also bears a close relationship to the theoretically and practically important spin Hall effect.\cite{DPshe71}  Indeed, the spin Hall current in the clean Rashba 2DEG arises as a transient in the process of building the Edelstein spin polarization.   

All of the theoretical studies mentioned in the previous paragraph were limited to the linear response regime -- weak electric field or weak spin injection --  meaning that the drift velocity of the electrons remains much smaller than the Fermi velocity and the non-equilibrium spin polarization is small.   There are good reasons for this choice, since this is in practice the regime in which virtually all of the experiments have been done.  The presence of impurity scattering limits the electron drift velocity to values much smaller than the Fermi velocity.

By contrast, in this paper we present a theoretical study of the Edelstein effect and its inverse in a {\it perfectly clean} Rashba 2DEG.   The absence of impurities allows the electrons to be accelerated to high velocities (comparable to the Fermi velocity) and thus to access the nonlinear regime.  There are several reasons for undertaking this study.  First of all, the model admits an elegant completely analytical solution, which is no common occurrence in this area of research.
Second, the solution is very instructive, bringing forth an unexpected connection with the classic Landau-Zener-Majorana model\cite{Landau32,Zener32,Majorana32} for the anti-crossing of two energy levels.
In brief, we find that when the drift velocity of the electrons reaches a sufficiently high value over a long time (i.e., for weak electric field) the Edelstein spin polarization (normally proportional to the electric field)  saturates to a limiting value corresponding to 100\% spin polarization of the electrons in an annulus of momentum space comprised between the Fermi momenta of the two Rashba bands.  However, if the acceleration is very high, the electron spins are unable to respond to a rapidly changing spin-orbit field, and the final polarization is much smaller than the saturation limit.  In addition, we find that, no matter how small the electric field is, a Landau-Zener anti crossing always occurs, for sufficiently large times, on part of the Fermi surface.  States on this part of the Fermi surface can either stay on the adiabatic track, or undergo a diabatic crossing, in which case their contribution to the final spin polarization is greatly suppressed.

In spite of the idealized character of our model, we believe that the results are of general interest, and some of our predictions could be tested in detail either in extremely clean electronic systems subject to strong electric fields or, possibly, in vapors of ultra cold fermonic atoms~\cite{Dalibard2011}.

This paper is organized as follows.  In Section II we introduce the model and set up the quantum kinetic equation for the response to an electric field.   In Section III we present the analytic solution of the kinetic equation and its long-time limit.  In Section IV we discuss the perturbative regime  $\gamma \ll 1$.  Section V presents the calculation of the transient spin Hall current.  Section VI describes the inverse (in the sense of Onsager reciprocity) of the nonlinear Edelstein effect.  Finally, section VII presents a qualitative discussion of the effects of disordered and the prospect for experimental observation.

\section{Model and kinetic equation}
Our model (two-dimensional electron gas with Rashba spin-orbit coupling $\alpha$) is described by the Hamiltonian
\be
H(\pv,t)=\frac{1}{2m}\left[\pv+e\Av(t)\right]^2 -\alpha([\pv+e\Av(t)] \times \sigmabold)\cdot \zv\,,
\ee 
where
\be
\Av(t)=-eEt \hat \xv
\ee
represents an electric field in the $x$ direction, and $t$ is time.  The system is assumed to be initially (i.e., at time $t=0$)  in equilibrium with a density matrix
\be
\rho(\pv,0)=f_+(p) {\bf 1}+f_-(p)(\hat \pv \times \sigmabold)\cdot \zv
\ee
where ${\bf 1}$ is the $2\times 2$ identity matrix,
\be
f_{\pm}(p)\equiv\frac{\theta(p_{F+}-p)\pm\theta(p_{F-}-p)}{2}
\ee
and $p_{F+}, p_{F-}$ ($p_{F+}<p_{F-}$)  are the Fermi momenta in the two chirality bands (see Figure 1) (we work at zero temperature for simplicity)
For $t>0$ the evolution of the density matrix is obtained by solving the  equation of motion
\be
\frac{\partial \rho(\pv,t)}{\partial t}+i[H(\pv,t), \rho(\pv,t)]=0
\ee

\begin{figure}\label{Populations}
\begin{center}
\includegraphics[width=2in]{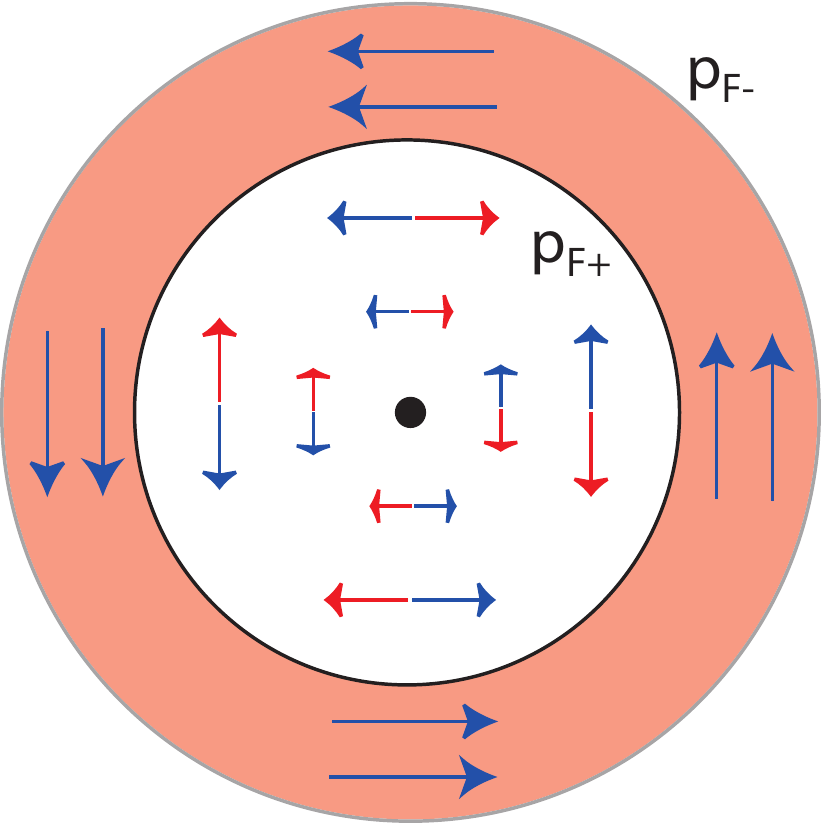}
\caption{Equilibrium distribution of electrons in momentum space.  The inner circle, of radius $p_{F+}$ has doubly occupied states.  The annulus $p_{F+}<p_F<p_{F-}$ (shaded area) has singly occupied states, with spins oriented along  the Rashba effective field.  The density of electrons in the annulus is $\sim 2n \alpha/v_F$, where $n$ is the total density: only these electrons respond to the application of the electric field.}
\end{center}
\end{figure}

We immediately observe that for $p<p_{F+}$, i.e. in the region where momentum states are ``doubly occupied" by electrons of opposite chiralities,  one has $f_-(p)=0$ and hence the initial distribution function is proportional to the identity matrix.  This commutes with the hamiltonian and therefore remains constant in time.  This means that the states with $p<p_{F+}$ constitute an inert background.  We will, from now on, focus exclusively on electrons in the annulus $p_{F+}<p<p_{F-}$ where the states are singly occupied.
In this region the initial density matrix describes a pure state in which the spin is aligned parallel to the Rashba field
\be
\Bv_R (\pv)= \alpha p \zv \times \hat\pv
\ee
The subsequent evolution of this  state is caused by the action of a time-dependent ``Edelstein field", which does not depend on $\pv$:
\be
\Bv_E(t) = \alpha \zv \times e\Av(t) = -\alpha (e Et) \hat\yv \,.
\ee
The problem is now reduced to calculating the evolution of a single spin, initially aligned along $\Bv_R(\pv)$ under the action of the total Zeeman field
\be
\Bv_{tot}(\pv,t)= \Bv_R (\pv)+\Bv_E(t) 
\ee
This problem is solved analytically in the next section. 
Once the spin dynamics is solved, we can also calculate the charge current response, which is the sum of a linearly varying ``diamagnetic" term due to the vector potential (this is simply the ballistic acceleration of the electrons) and the spin term $\alpha S^y$. 

\section{Analytic solution of the model}
\begin{figure}\label{CoordinateSystem}
\begin{center}
\includegraphics[width=3in]{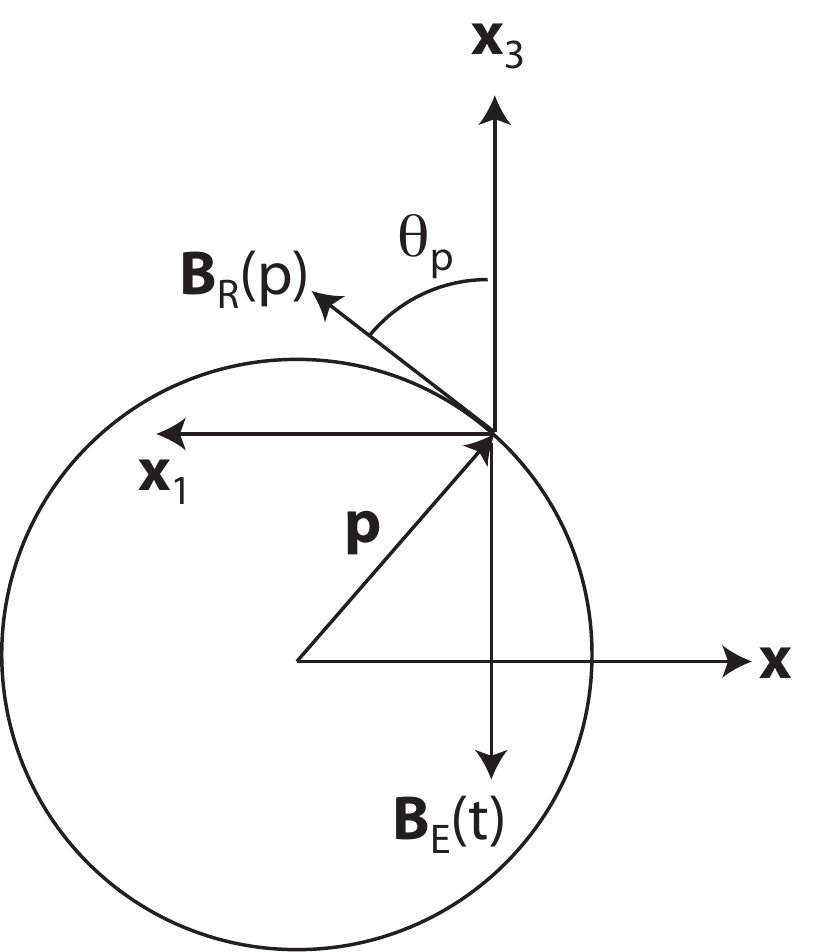}
\caption{Coordinate system used in the calculations.  The spin is initially oriented along the Rashba field $\Bv_R(\pv)$. The angle $\theta_\pv$ between the Rashba field and the $\xv_3$ axis is the same as the angle between $\pv$ and the $\xv$ axis.  $\Bv_E(t)$ is the Edelstein field, which grows linearly in time, starting from $0$ at the initial time.}
\end{center}
\end{figure}
Our system of coordinates is shown in Fig.~2.
The fixed direction of the Edelstein field ($\hat \yv$) is taken as our $\hat\xv_3$ axis.  The original $-\hat\xv$ and $\zv$ directions become our $\hat \xv_1$ and $\hat \xv_2$ respectively.  We denote by $\theta_\pv$ the angle between $\hat \pv$ and the standard $\xv$ axis.  Thus the Rashba field forms an angle $\theta_\pv$  with the $\xv_3$ axis and we can write
\ber
\Bv_R(\pv)&=&\alpha p(\cos\theta_\pv \hat\xv_3+\sin\theta_\pv \hat\xv_1)\,,\nn\\
\Bv_E(t) &=& -\alpha eEt \hat\xv_3\,,\nn\\
\Bv_{tot}(\pv,t)&=&(\alpha p\cos\theta_\pv-\alpha eEt )\hat\xv_3+\alpha p \sin\theta_\pv \hat\xv_1\,,
\eer
We use the projections of the spin along the $\xv_3$ axis as the basis for our representation of the spin.
Then the initial state of the spin of the electron with momentum $\pv$ (in the range $p_{F+}<p<p_{F-}$)
is
\be
|\psi_\pv(0)\rangle = \cos\frac{\theta_\pv}{2}|\up\rangle+\sin\frac{\theta_\pv}{2}|\down\rangle\,.
\ee
The time-dependent hamiltonian is
\be
\hat H_\pv(t) = \left(\begin{array}{cc}
-\alpha p \cos \theta_\pv+\alpha eEt  &-\alpha p \sin\theta_\pv\\
-\alpha p \sin\theta_\pv& +\alpha p \cos\theta_\pv-\alpha eEt 
\end{array}\right)\,,
\ee
which is recognized to be the canonical Landau-Zener Hamiltonian\cite{Landau32,Zener32}, which is often used to describe the  transition probability between two energy levels ($-\alpha p \cos \theta_\pv +\alpha eEt$ and  $\alpha p \cos \theta_\pv -\alpha eEt$ in this case),  which anticross  as a function of time (this problem was first studied in the context of spin physics by E. Majorana~\cite{Majorana32}.  For a pedagogical discussion of the model see Ref.~\onlinecite{Wittig05}.)
The dynamics depends crucially on whether the rate of variation  of the energy levels as they cross, $e E/(p\cos\theta_\pv)$,  is small or large compared to the magnitude of the matrix element that couples the two levels,  $\alpha p \sin\theta_\pv$.  In the former case the process is adiabatic and the spin follows faithfully the magnetic field; in the latter the spin has no time to respond to the rapidly changing conditions and remains close to its initial orientation.
An overall measure of non-adiabaticity is therefore given by the ratio  
\be
\gamma \equiv \frac{eE}{\alpha p_F^2}=\frac{eEL_s}{E_F}\,,
\ee
where $L_s = 1/(2 m\alpha)$ is the spin precession length (here and in the following we set $\hbar=1$).  
The adiabatic regime is characterized by $\gamma\ll 1$ and the non-adiabatic one by $\gamma \gg 1$.  We stress that $\gamma$  is only an {\it average} indicator of adiabaticity: states with $\theta_\pv =0$ will never be adiabatic, no matter how small $\gamma$ is, because the matrix element coupling the two levels vanishes when $\theta_\pv=0$.
 
In order to simplify the calculations that follow we  express time in units of $(\alpha p_F \sqrt{\gamma})^{-1}$: 
\be
t \equiv \frac{\tau}{\alpha p_F\sqrt{\gamma}}\,.
\ee
Then the Hamiltonian (also expressed in units of  $\alpha p_F \sqrt{\gamma}$) takes the form
\be
\hat H_\pv(\tau) = \left(\begin{array}{cc}
\tau  -\tau_\pv &- \Delta_\pv\\
-\Delta_\pv&  -(\tau  -\tau_\pv )
\end{array}\right)\,,
\ee
where we have introduced the notation
\be\label{Deltap}
\tau_\pv \equiv \frac{\cos\theta_\pv}{\sqrt{\gamma}}\,,~~~~~\Delta_\pv \equiv \frac{\sin\theta_\pv}{\sqrt{\gamma}}\,.
\ee  
Notice that $\tau_\pv$ is the (dimensionless) time for which the gap between the levels would close in the absence of the coupling $\Delta_\pv$.  The absolute value of $\Delta_\pv$ is the ``residual gap" at the anti crossing point.   Because the hamiltonian depends on time only via the combination $\tau-\tau_\pv$ it is evident that we can represent the  solution of the time-dependent Schr\"odinger equation in the form
\be
|\psi_\pv(\tau)\rangle = u_\pv(\tau-\tau_\pv)|\up\rangle+v_\pv(\tau-\tau_\pv)|\down\rangle\,,
\ee
where the amplitudes $u_\pv(\tau)$ and $v_\pv(\tau)$ satisfy the system of equations
\ber\label{Equations}
i\dot u_\pv(\tau)&=&\tau u_\pv(\tau)- \Delta_\pv v_\pv(\tau)\nn\\
i\dot v_\pv(\tau)&=&-\tau v_\pv(\tau)- \Delta_\pv u_\pv(\tau)\,.
\eer
where the dot denotes the derivative with respect to $\tau$ and the initial conditions are
\be\label{InitialConditions}
u_\pv(-\tau_\pv)=\cos(\theta_\pv/2)\,,~~~~~~~~v_\pv(-\tau_\pv)=\sin(\theta_\pv/2)\,.
\ee
Two mutually orthogonal solutions of Eqs.~(\ref{Equations}) are readily found in terms of 
parabolic cylinder functions $D(\nu,z)$  (see Appendix for details of the derivation) as follows:
\ber\label{Sol1}
u_\pv^{(1)}(\tau)&=&D(-i\Delta_\pv^2/2, e^{-i\frac{3\pi}{4}}\sqrt{2}\tau) e^{-\pi \Delta_\pv^2/8}\nn\\
v_\pv^{(1)}(\tau)&=&e^{i\frac{\pi}{4}} (\Delta_\pv/\sqrt{2})  D(-i\Delta_\pv^2/2-1, e^{-i\frac{3\pi}{4}}\sqrt{2}\tau) e^{-\pi \Delta_\pv^2/8}\,,
\eer
and
\ber \label{Sol2}
u_\pv^{(2)}(\tau)&=&-[v_\pv^{(1)}(\tau)]^*\nn\\
v_\pv^{(2)}(\tau)&=&[u_\pv^{(1)}(\tau)]^*\,.
\eer
The constant $e^{-\pi \Delta_\pv^2/8}$ has been chosen so that both solutions satisfy the normalization condition 
\be
|u_\pv(\tau)|^2+|v_\pv(\tau)|^2=1\,,
\ee
as they should.  It turns out that these solutions are precisely the solutions of the classic Landau-Zener problem, in which the system is prepared at $\tau=-\infty$  in one the two eigenstates $|\up\rangle$ or $|\down\rangle$.  Indeed, making use of the asymptotic behavior of the parabolic cylinder functions \cite{Bateman-Erdelyi_V2,Gradshtein} we find
\be\label{Asymptotics-u}
 u_\pv^{(1)}(\tau) =e^{-i[\tau^2/2+\Delta_\pv^2 \ln (\sqrt{2}\tau)/2]}\left\{\begin{array}{c}1~~~~~~~~~~~~(\tau \to -\infty)\\e^{-\pi\Delta_\pv^2/2}~~~~~~~(\tau \to  +\infty) \end{array}\right. 
 \ee
 and
 \be \label{Asymptotics-v}
  v_\pv^{(1)}(\tau) =e^{i[\tau^2/2+\Delta_\pv^2 \ln (\sqrt{2}\tau)/2+\pi/4+\arg \Gamma(1-i\Delta_p^2/2)]}\left\{\begin{array}{c}0~~~~~~~~~~~~~(\tau \to -\infty)\\ \sqrt{1-e^{-\pi\Delta_\pv^2}}~~~(\tau \to  +\infty) \end{array}\right.
 \ee 
where $\Gamma(z)$ is the gamma function. This leads us to the identification of
 \be
 P^{LZ}_\pv \equiv  e^{-\pi\Delta_\pv^2}=e^{-\frac{\pi}{\gamma}\sin^2\theta_\pv}
 \ee
 as the survival probability of the initial level, i.e., the probability of staying on the {\it diabatic} track at the anticrossing point).  $1-P^{LZ}_\pv$ is the Landau-Zener transition probability, i.e. the probability of avoiding the crossing.   From the definition~(\ref{Deltap}) of $\Delta_\pv$ we see that 
 $P^{LZ}_\pv$ tends to $0$ for $\gamma \to 0$ and to $1$ for $\gamma \to \infty$, unless $\theta_\pv=0$, in which case it is $1$ for all values of $\gamma$ (this indicates the complete breakdown of the adiabatic approximation).
 
 Finally, we  seek our solution as a linear combination of the two orthogonal solutions~(\ref{Sol1}) and~(\ref{Sol2}) that satisfy the initial conditions of our problem, as stated in Eq.~(\ref{InitialConditions}). 
 We set
 \ber
 u_\pv(\tau-\tau_\pv)&=&A_\pv u^{(1)}_\pv(\tau-\tau_\pv)+B_\pv u^{(2)}_\pv(\tau-\tau_\pv)\nn\\
 v_\pv(\tau-\tau_\pv)&=&A_\pv v^{(1)}_\pv(\tau-\tau_\pv)+B_\pv v^{(2)}_\pv(\tau-\tau_\pv)\,,
 \eer
 and determine $A$ and $B$ from the conditions
 \ber
 \cos \frac{\theta_\pv}{2}&=& A_\pv u^{(1)}_\pv(-\tau_\pv)+B_\pv  u^{(2)}_\pv(-\tau_\pv)\nn\\
 \sin \frac{\theta_\pv}{2}&=& A_\pv v^{(1)}_\pv(-\tau_\pv)+B_\pv v^{(2)}_\pv(-\tau_\pv)\,.
 \eer
 On account of the relation~(\ref{Sol2}) between the $(1)$ and $(2)$  solutions we easily find
 \ber
 A_\pv &=& \cos \frac{\theta_\pv}{2}[u^{(1)}_\pv(-\tau_\pv)]^*+\sin \frac{\theta_\pv}{2}[v^{(1)}_\pv(-\tau_\pv)]^*\nn\\
 B_\pv &=&\sin \frac{\theta_\pv}{2}u^{(1)}_\pv(-\tau_\pv)-\cos \frac{\theta_\pv}{2}v^{(1)}_\pv(-\tau_\pv)\,.
 \eer
This, combined with the expressions of Eq.~(\ref{Sol1}) completes the analytical solution of our model. 

\begin{figure}
\begin{center}\label{ResolvedTimeEvolution}
\includegraphics[width=4in]{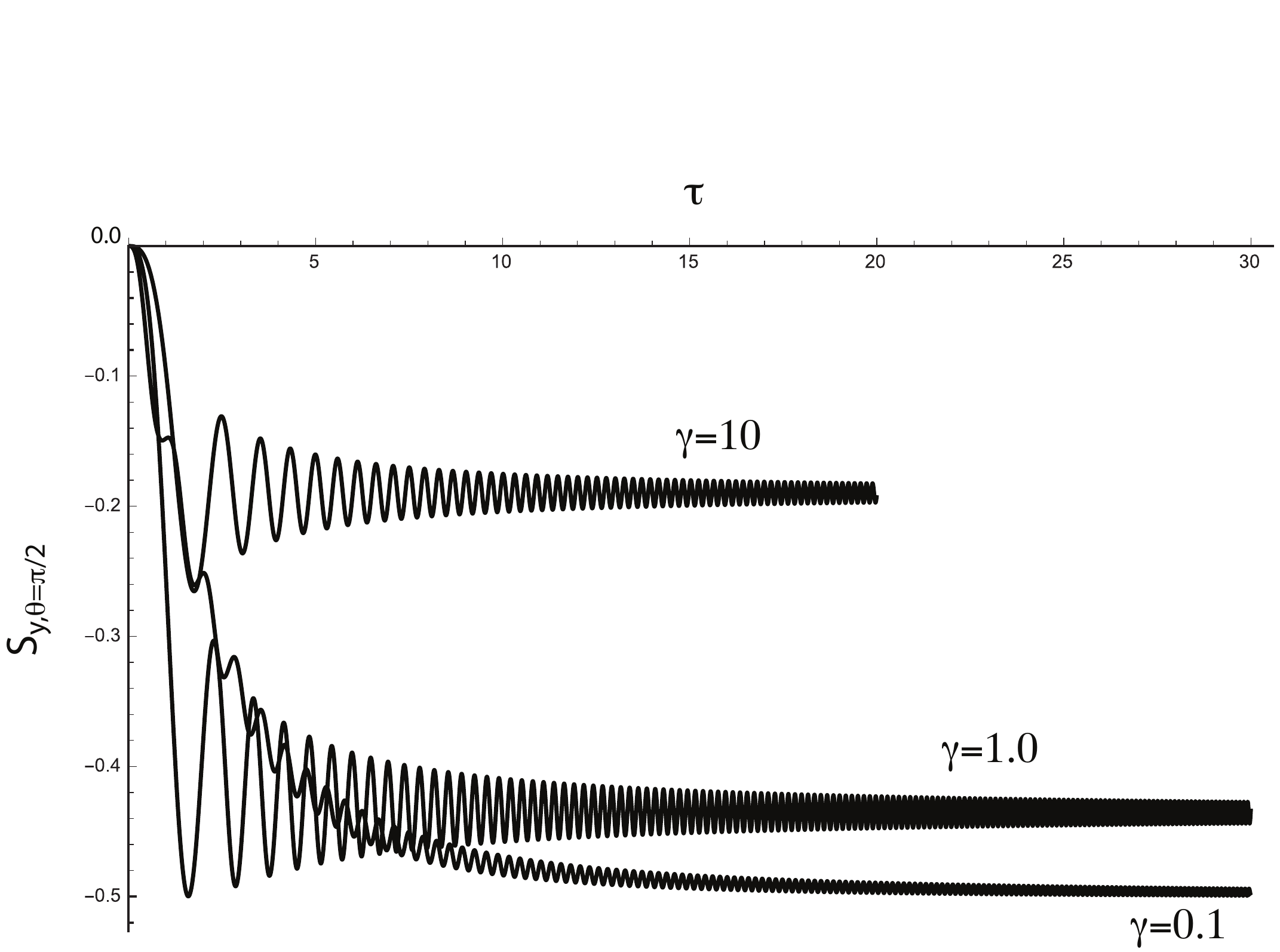}
\caption{Plot of the Time evolution of the most responsive spin ($\theta_\pv=\pi/2$)  vs $\tau$ for three different values of $\gamma$.  Apart from stronger oscillations, which are washed out by angular integration, these plots are qualitatively similar to the plots for the time evolution of the total spin polarization in Fig.~4.}
\end{center}
\end{figure}

\begin{figure}
\begin{center}\label{FullTimeEvolution}
\includegraphics[width=4in]{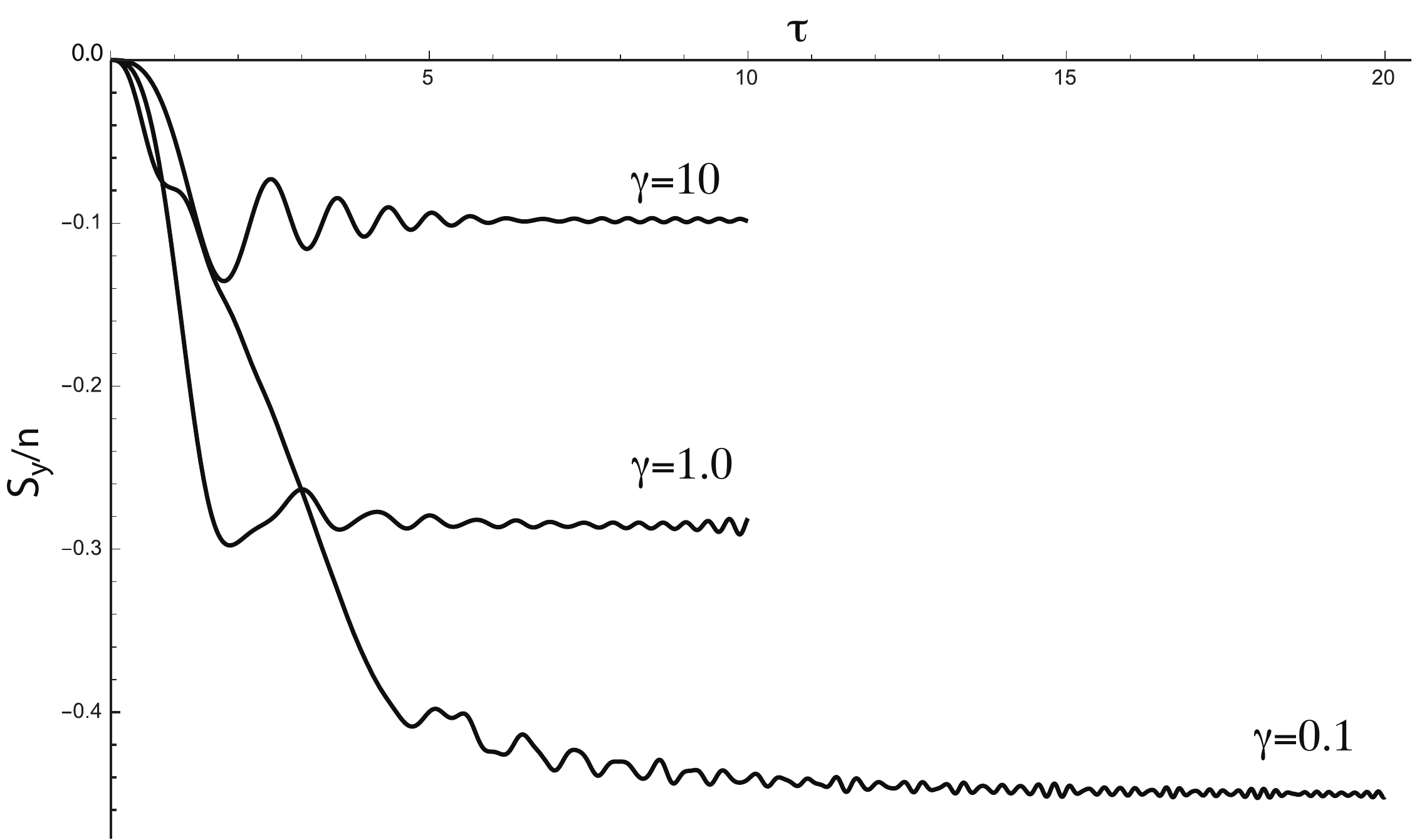}
\caption{Plot of the Edelstein spin polarization $S_y/n$ in units of $\alpha/v_F$ vs $\tau$ ($\gamma \tau=-v(t)/v_F$) for three different values of $\gamma$.  Notice that long-time limits are given by the asymptotic formula~(\ref{LongTimeLimit}) and are equal to  -0.449323, -0.286822, -0.0982924 for $\gamma$=0.1,1, and 10 respectively. }
\end{center}
\end{figure}

In terms of the amplitude $u_\pv(\tau-\tau_\pv)$ we express the $y$-component of the spin  as
\ber\label{Edelstein1}
S_{\pv,y}(\tau)&=&\frac{1}{2} \left(|u_\pv(\tau-\tau_\pv)|^2-|v_\pv(\tau-\tau_\pv)|^2\right)\nn\\
&=&|u_\pv(\tau-\tau_\pv)|^2-\frac{1}{2}\,.
\eer
%The deviation from the initial equilibrium state is
%\ber
%\delta S_{\pv,y}(\tau)&=&|u_\pv(\tau-\tau_\pv)|^2-\frac{1}{2}-\frac{1}{2}\cos\theta_\pv \nn\\
%&=&|u_\pv(\tau-\tau_\pv)|^2-\cos^2\frac{\theta_\pv}{2}\,.
%\eer
At last, the total Edelstein spin polarization is the number of states satisfying $p_{F+}<p<p_{F-}$ (i.e., $2\alpha n/v_F$)  times the angular average of $S_{\pv,3}(\tau)$:
\be\label{Edelstein2}
S_y(\tau)= \frac{2\alpha n}{v_F} \frac{1}{2\pi}\int_0^{2\pi}d\theta_\pv \left(|u_\pv(\tau-\tau_\pv)|^2-\frac{1}{2}\right)\,,
\ee 
where the magnitude of $\pv$ is approximated as $p_F$.  

Figs. 3 and 4 present plots of the exactly calculated analytical solution as a function of time for three representative values of $\gamma$: $\gamma=0.1$ (quasi-adiabatic regime), $\gamma=1$ (intermediate regime), and $\gamma=10$ (sudden switch-on, or ``anti-adiabatic" regime).  Fig. 3 shows the time evolution of the most significant (i.e., most responsive) spins with $\theta_\pv=\pi/2$. Fig. 4 shows  the total spin response integrated over the Fermi surface.

Let us now consider the long-time limit of the solution.  Substituting the asymptotic forms~(\ref{Asymptotics-u}) and (\ref{Asymptotics-v}) into the expression~(\ref{Edelstein1})  for the Edelstein spin polarization we obtain
\ber\label{LongTimeLimit}
 |u_\pv(\infty)|^2&=&P^{LZ}_\pv  |A_\pv|^2+ (1-P^{LZ}_\pv)|B_\pv|^2\nn\\
 &-&2 \sqrt{P^{LZ}_\pv (1-P^{LZ}_\pv)}~\Re e \left\{ A_\pv B_\pv^* e^{i \left[\frac{\pi}{4}+\arg \Gamma \left(1-i\frac{\Delta_\pv^2}{2}\right)\right]} \right\}\,.
\eer

\begin{figure}
\begin{center}\label{InfiniteTime}
\includegraphics[width=4in]{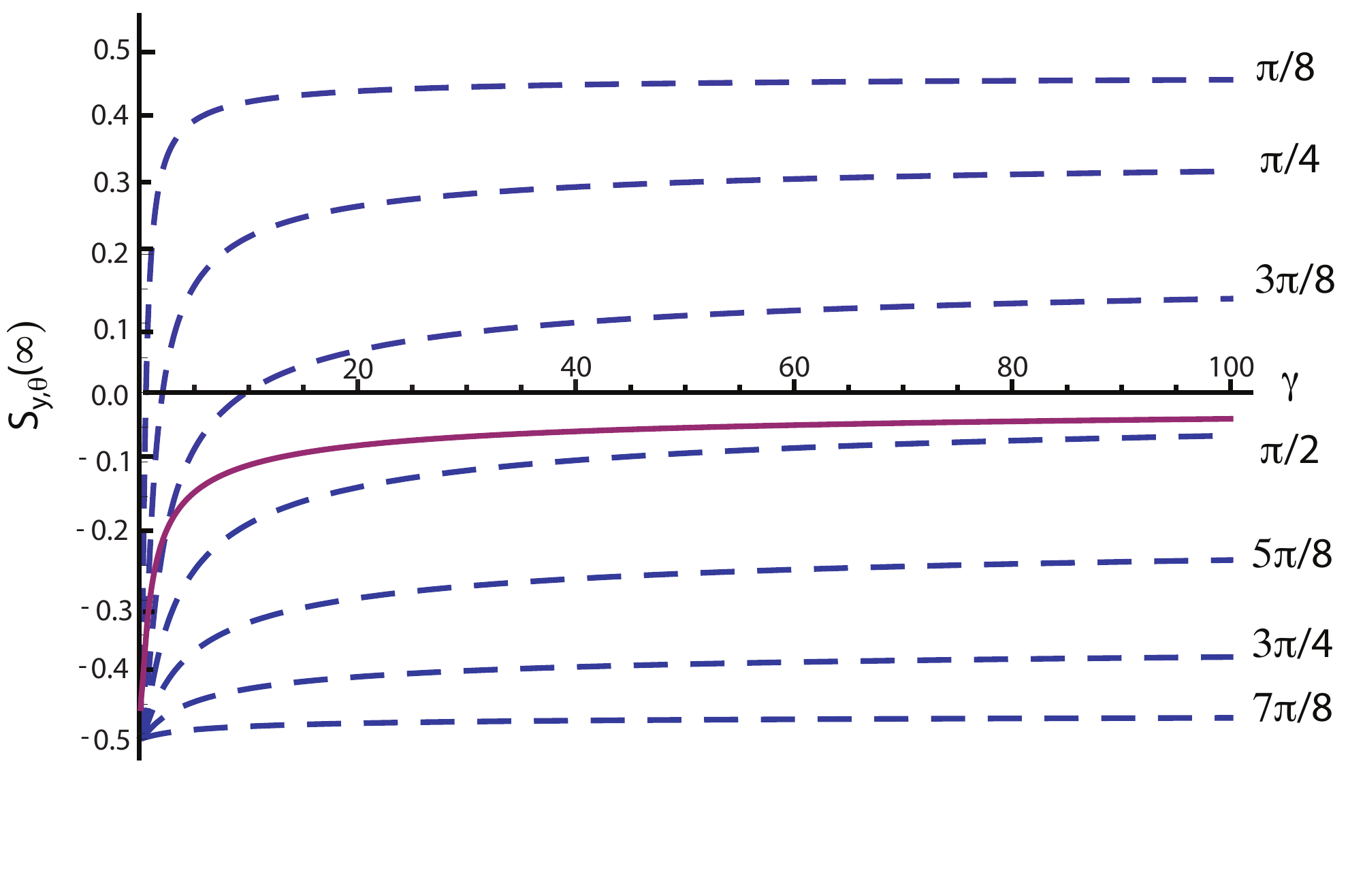}
\caption{Plot of $S_{y,\pv}(\infty)$ vs. $\gamma$ for various values of $\theta_\pv$ ranging from $\pi/8$ (top dashed line) to $7\pi/8$ (bottom dashed line) in steps of $\pi/8$.  In all cases, the limiting value at $\gamma=0$ is $S_{y,\pv}(\infty)=-1/2$, signifying perfectly adiabatic response.  The limiting value for $\gamma \to \infty$ is $\cos \theta_\pv$, signifying that  $S_{y,\pv}(\infty)=S_{y,\pv}(0)$.  The solid line is the angular average of $S_{y,\pv}(\infty)$.   Notice that the transition from the adiabatic to the non adiabatic regime occurs at lower values of $\gamma$ when $\theta_\pv$ is close to $0$ and $\pi$ (top and bottom lines).  In fact, the adiabatic regime is never attained at $\theta_\pv=0$ or $\pi$.  }
\end{center}
\end{figure}

This remarkable result tells us that the Edelstein spin $S_{y,\pv}$ tends, for large time, to a constant limiting value, $|u_\pv(\infty)|^2-1/2$.  This is expected, because in this limit the Edelstein field is much larger than the original Rashba field, and the projection of the spin along its direction becomes essentially a constant of the motion.  However, the limiting value  is strongly dependent on $\gamma$. For $\gamma \ll 1$ the evolution of the spin is generally adiabatic, with the exception of states with $\theta_\pv \sim 0$  for which $\sin^2\theta_\pv/\gamma <1$ (see discussion in the next section).  Thus, with the exception of $\theta_\pv$ in the immediate vicinity of $0$, the  spin follows the orientation of the total effective magnetic field, settling in a state with $S_{y,\pv} =-1/2$ for $\tau\to\infty$.  Mathematically, this corresponds to the fact that $|u_\pv(\infty)|^2 \to 0$ for $\gamma \to 0$.

In the opposite limit of $\gamma \gg 1$ the evolution of the spin is strongly non-adiabatic.  Basically the projection of the spin along the direction of the Edelstein field does not have enough time to change: it remains equal to the initial value in the limit of $\gamma \to \infty$.  Mathematically this is expressed by the fact that  $|u_\pv(\infty)|^2 \to \cos^2\frac{\theta_\pv}{2}$ for $\gamma \to \infty$.   Fig. 5 shows the infinite time limit of $S_{y,\pv}(\infty)=|u_\pv(\infty)|^2- \frac{1}{2}$ as a function of $\gamma$.

\section{Limit of $\gamma \ll 1$}
Let us examine more closely the important limit of $\gamma \ll 1$, i.e., weak electric field.  For a given angle, $\theta_\pv$, the  parameter that controls the ``adiabaticity" of the dynamics is the ratio of the fractional rate of change of the effective field $|\dot \Bv_\pv|/|\Bv_\pv|$ to the energy difference between the two opposite orientations of the spin in the total Zeeman field, $|\Bv_\pv|$.  This gives
\ber\label{AdiabaticCondition}
\eta_\pv=\frac{|\dot \Bv_\pv|}{B_\pv^2} &=&\frac{\alpha e E}{(\alpha p_F \cos\theta_\pv-\alpha e Et)^2+(\alpha p_F\sin\theta_\pv)^2}\nn\\
&=&\frac{\gamma}{(\cos\theta_\pv-\sqrt{\gamma}\tau)^2+\sin^2\theta_\pv}\,.
%\nn &=&\frac{\gamma}{1+(\gamma\tau)^2-2\gamma\tau\cos\theta_\pv}\,.
\eer
The adiabatic regime occurs when $\eta_\pv \ll 1$.  For small $\gamma$, this will always be the case for the states with  $\pi/2<\theta_\pv<\pi$ for in this case the denominator is always larger than $1$.     On the other hand,  for states with  $0<\theta_\pv<\pi/2$, the denominator reaches the minimum value  $\sin^2\theta_\pv$ when $\tau = \cos\theta_\pv/\sqrt{\gamma}=\tau_\pv$.   The condition of adiabaticity is satisfied only for  $\gamma \ll \sin^2\theta_\pv$ or, for a given $\gamma$, when 
\be
|\sin\theta_\pv|\gg \sqrt{\gamma}\,.
\ee
Thus, the adiabatic approximation always fails for $\theta_\pv$ close to $0$, no matter how small $\gamma$ is. The non-adiabaticity ``kicks in" at $\tau=\tau_\pv$, i.e., at the crossing of the levels: this occurs when the velocity of the electrons equals the Fermi velocity.   

These qualitative considerations are confirmed by an explicit calculation of the adiabatic spin response to the Edelstein field.  We have
\be
u^{ad}_\pv (\tau)=\frac{1}{2}\left(1-\frac{\tau}{\sqrt{\tau^2+\Delta_\pv^2}}\right)^{1/2}
\ee
and therefore the $y$-component of the spin is given, according to Eq.~(\ref{Edelstein1}) by
\be
S^{ad}_{\pv,y}(\tau) = \frac{1}{2} \frac{\sqrt\gamma \tau -\cos\theta_\pv}{\sqrt {1+\gamma\tau ^2-2\sqrt{\gamma} \tau \cos\theta_\pv}}
%&=&\frac{1}{2}\frac{(\alpha p \cos\theta_\pv -\alpha eEt )\hat\xv_3+\alpha p \sin\theta_\pv \hat\xv_1}{\sqrt{(\alpha p \cos\theta_\pv-\alpha eEt )^2+(\alpha p \sin\theta_\pv)^2}}\nn\\
%&=&\frac{1}{2}\frac{(\cos\theta_\pv-\gamma \tau )\hat\xv_3+\sin\theta_\pv \hat\xv_1}{\sqrt{1+(\gamma\tau )^2-2\cos\theta_\pv\gamma \tau}}\,.
\ee
where we have made use of Eqs.~(\ref{Deltap}) for $\tau_\pv$ and $\Delta_\pv$.  
The Edelstein spin density is then obtained by performing an elementary integration over the angle $\theta_\pv$:
\be\label{SyIntegral}
S^{ad}_y(\tau)= \frac{\alpha n}{v_F}\frac{1}{2} \frac{1}{2\pi}\int_0^{2\pi}d\theta
\frac{\cos\theta -\sqrt{\gamma} \tau }{\sqrt{1+(\sqrt{\gamma}\tau )^2-2 \sqrt{\gamma} \tau \cos \theta}}
\ee
This integral is elementary and gives
\be\label{SyAdiabatic}
S^{ad}_y(\tau)=- \frac{\alpha n}{v_F}\frac{1}{2}\left\{\frac{(\sqrt{\gamma}\tau+1)E\left[\frac{4\sqrt{\gamma}\tau}{(\sqrt{\gamma}\tau+1)^2}\right]+(\gamma\tau-1)K\left[\frac{4\sqrt{\gamma}\tau}{(\sqrt{\gamma}\tau+1)^2}\right]}{\pi\sqrt{\gamma}\tau}\right\}
\ee
where $E$ and $K$ are the standard elliptic integrals \cite{Gradshtein}.
\begin{figure}
\begin{center}
\includegraphics[width=4in]{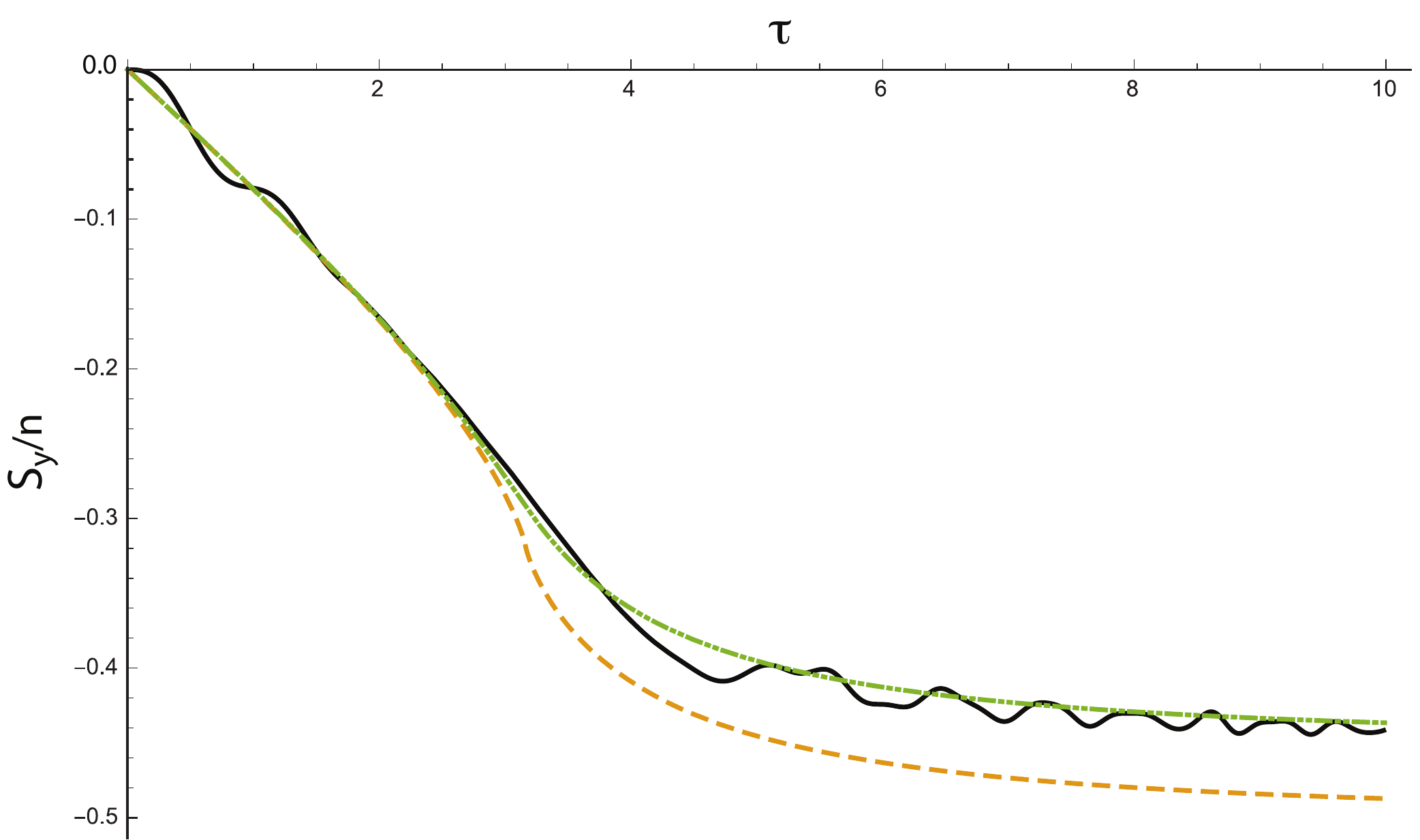}
\caption{Plot of the Edelstein spin polarization $S_y/n$ in units of $\alpha/v_F$ vs $\tau$ for $\gamma=0.1$, taken as representative of the ``adiabatic regime". The dashed line is  the adiabatic approximation of Eq.~(\ref{SyAdiabatic}).   The solid line  is the exact solution.  The discrepancy between the two curves for $\gamma \tau>1$ is attributed to the persistent non-adiabatic response of spins with $\theta_\pv \sim0$ and $\pi$. The dash-dotted line is calculated by the improved formula~(\ref{SyCorrected}) and it is seen to be in much better agreement  with the exact results.}
\end{center}
\end{figure}

Fig. 6 shows the Edelstein spin polarization as a function of $\sqrt{\gamma} \tau=|v(\tau)|/v_F$, where $v(\tau) = \frac{eE}{m}\frac{\tau}{\sqrt{\gamma}\alpha p_F}$ is the velocity of the freely accelerating electrons at time $t = \frac{\tau}{\sqrt{\gamma}\alpha p_F}$.
In the linear response regime $|v(t)|/v_F \ll 1$ this formula reduces to
\be\label{LinearEdelstein}
S_y(t)\simeq\frac{\alpha n}{2v_F} \frac{eEt}{mv_F} = \frac{N_0}{2} \alpha eE t\,,
\ee
where $N_0=n/\epsilon_F=m/(2\pi)$ is the density of states (per spin).  This is the standard formula for the linear Edelstein effect.   Notice that the shortcomings of the adiabatic approximation do not show up in this regime, because one is never close to the Landau-Zener anticrossing.
When $\sqrt{\gamma} \tau=1$, (meaning that $|v(\tau)|=v_F$)  a non-analyticity (logarithmically infinite derivative) is present and clearly visible in the plot of Fig. 7. In the next section we show that this leads to an unphysical divergence of the spin current.  These are all artifacts of the adiabatic approximation, and can be cured in an approximate but very effective manner by multiplying the integrand of Eq.~(\ref{SyIntegral}) by the probability $1-P^{LZ}_\pv$ of the Landau-Zener transition, i.e., the probability of actually staying on the adiabatic track.  
The resulting formula,
\be\label{SyCorrected}
S_y(\tau)= \frac{\alpha n}{v_F}\frac{1}{2} \frac{1}{2\pi}\int_0^{2\pi}d\theta
\frac{\cos\theta -\sqrt{\gamma} \tau }{\sqrt{1+(\sqrt{\gamma}\tau )^2-2 \sqrt{\gamma} \tau \cos \theta}}
\left(1-e^{-\frac{\pi}{\gamma}\sin^2\theta}\right)\,,
\ee 
is numerically evaluated and plotted  in Fig. 6.
This formula is free of pathological behaviors at the anticrossing point, since the contribution from the ``unresponsive" states with $\sin \theta_\pv <\sqrt{\gamma}$ has been suppressed by the Landau-Zener transition probability. 

\section{Spin Hall Current}
In this section we calculate the transient spin Hall current, which accompanies the electric current.
From the Heisenberg equation of motion
\be
\frac{dS_y(t)}{dt}=i[H(t),S_y]
\ee
we immediately obtain
\be
\frac{dS_y(t)}{dt}=-\alpha p_y\sigma_z =  -2 m\alpha J^z_y
\ee
where $J^z_y\equiv \frac{p_y}{2m}\sigma_z$ is the operator of the spin current.  The total spin current density is therefore proportional to the time derivative of the total Edelstein spin density.  The spin current can be calculated analytically, according to the formula
\be
J^z_y(\tau)=-\frac{1}{2m\alpha}\frac{dS_y(t)}{dt}=-\frac{v_F\sqrt{\gamma}}{2}\dot S_y(\tau)\,.
%=-\frac{2\sqrt{\gamma}\alpha n}{\pi}\int_0^\pi d\theta_\pv \Re e[u^*_\pv(\tau-\tau_\pv)\dot u_\pv(\tau-\tau_\pv)]\,, 
\ee
%where we have made use of the fact that  $\frac{dS_y(t)}{dt} = \alpha p_F\sqrt{\gamma}\dot S_y(\tau)$.
In the linear response regime, making use of Eq.~(\ref{LinearEdelstein}) we recover the well-known result
\be
J^z_y = -\frac{e}{8 \pi} E\,.
\ee

\begin{figure}
\begin{center}
\includegraphics[width=3in]{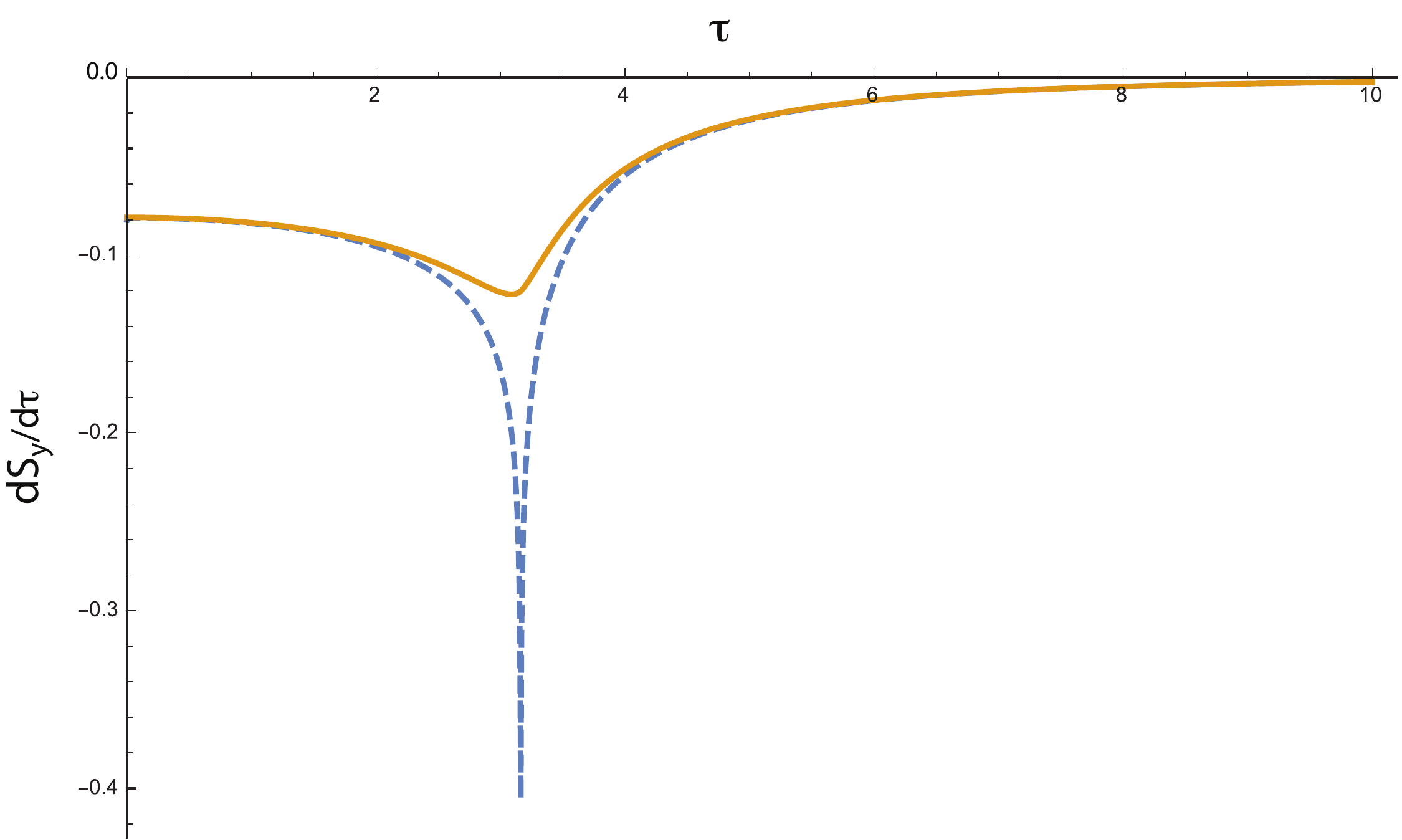}
\caption{Dashed line: plot of the transient spin Hall current $j^z_y \propto dS_y/d\tau$ in units of $-eE/(4\pi)$ vs $\tau$ calculated according to the adiabatic formula~(\ref{SyAdiabatic}) for $\gamma=0.1$: the unphysical divergence occurs when $\sqrt{\gamma}\tau=1$, that is to say when the drift velocity of the electrons equals the Fermi velocity.  The solid line is calculated according to the improved formula~(\ref{SyCorrected}) and it is free of singularities.}
\end{center}
\end{figure}
Beyond the linear response regime (but still in the quasi-adiabatic regime $\gamma\ll 1$) we can calculate $j_y^z(t)$ from the time derivative of Eq.~(\ref{SyCorrected}), or from the exact formulas.  The results are plotted in Fig. 7.

\section{Inverse Edelstein effect}
The calculations we have done in the previous sections can be straightforwardly adapted to the inverse Edelstein effect.  This is the reciprocal, in the Onsager sense, of the direct effect.  We apply a magnetic field that couples to the $y$ component of the spin and varies linearly in time:
\be
B_y(t)= \Gamma t\,,
\ee
and we calculate the charge current 
\be
J_x =  -e\sum_n\left \{\frac{p_{n,x}}{m}-\alpha \sigma_{n,y}\right\}
\ee 
that flows in response.  The sum runs over the particles, labelled by $n$.  Since the momentum distribution is not affected by $B_y(t)$ we immediately conclude that the expectation value of  $\sum_n\frac{p_{n,x}}{m}$ vanishes and we are left with
\be
J_x(t)=2\alpha e S_y (t)\,.
\ee 
Observe that $B_y(t)/2$  corresponds to  $-\alpha e A_x(t)$ of our previous calculation, and the ``spin injection field" $\Gamma/2$ corresponds to $\alpha e E$.   Thus, if we represent the result of our calculation for the direct Edelstein effect in the form
\be
S_y(t) = F\left(\frac{eE}{\alpha p_F^2},\frac{e E t}{p_F}\right)\,,
\ee
where we have used the fact that $\gamma = eE/(\alpha p_F^2)$ and  $\sqrt{\gamma} \tau= eEt/p_F$ and $F$ is the appropriate function of the two arguments, we immediately conclude that the current generated by the inverse Edelstein effect is
\be
J_x(t) = 2 \alpha e  F\left(\frac{\Gamma}{2\alpha^2 p_F^2},\frac{\Gamma t}{2\alpha p_F}\right)\,.
\ee
In particular, in the limit of large times the current tends to a limiting value proportional to $\alpha^2$.  

It is interesting to observe that the direct and inverse  ``conductivities"  are
\be
\frac{\partial S_y(t)}{\partial E} = \frac{e}{\alpha p_F^2} F^{(1)}\left(\frac{eE}{\alpha p_F^2},\frac{e E t}{p_F}\right)+ \frac{e t}{p_F} F^{(2)}\left(\frac{eE}{\alpha p_F^2},\frac{e E t}{p_F}\right)\,,
\ee
and
\be
\frac{\partial J_x(t)}{\partial \Gamma} = \frac{e}{\alpha p_F^2} F^{(1)}\left(\frac{\Gamma}{2\alpha^2 p_F^2},\frac{\Gamma t}{2\alpha p_F}\right)+ \frac{e t}{p_F} F^{(2)}\left(\frac{\Gamma}{2\alpha^2 p_F^2},\frac{\Gamma t}{2\alpha p_F}\right)\,,
\ee
where $F^{(1)}$  and  $F^{(2)}$ denote the partial derivatives of $F$ with respect to the first and the second argument respectively.  
This generalized Onsager reciprocity relation remains valid well beyond the linear response regime. The conductivities of the direct and inverse processes are identical if and only if they are evaluated at fields  $E$ and $\Gamma$ that satisfy the reciprocity condition  $2\alpha e E= \Gamma$.  This condition is of course satisfied in zero field,  where our relation reduces to the standard reciprocity relation of linear response theory.

\section{Effect of disorder and prospects for observation}
Up to this point we have completely neglected the effect of impurity scattering.  This has allowed us to obtain an exact and completely analytical solution.  However, it raises questions about the possibility of observing the nonlinear effect in realistic system. Cold trapped atoms, being intrinsically free of disorder, could provide an opportunity to do this. The challenge is to find a way to create an artificial Rashba spin-orbit field for cold atoms.  So far only pure gauge fields (e.g., the equal weight combination of Rashba and Dresselhaus fields) have been successfully synthesized by exposing the atoms to multiple laser fields which induces a quantum coherence between two hyperfine levels of the atom (the ``spin" degree of freedom).\cite{Lin2009,Lin2011,Dalibard2011}  However, there seems to be no obstacle, in principle, to the realization of an artificial two-dimensional Rashba field, and, in fact, theoretical proposals to this effect have already been put forward.\cite{Sau2011,Xu2012}

%{\bf (There is no Edelstein effect for the pure gauge, at least in the usual setting in the homogeneous system, because the pure gauge SOC can be completely gauged away from the equations. Alternatively: in this case the spin projection on the direction of the SO field is conserved and therefore the spin part of the wave functions is not changing. We can also interpret this as complete absence of the gap at the crossing for all states. Hence all states are 100\% nonadiabatic.)}

Let us further consider the case of electrons in clean systems.  Due to the unavoidable presence of impurities momentum is not conserved and the distribution function is no longer constant in momentum space.  In the relaxation-time approximation it shifts along the direction of the electric field by a time dependent quantity $mv(t)$ which eventually saturates to the Drude value $m v(t)=-e E \tau$, where $\tau$ is the electron-impurity scattering time. 
%{\bf (I would like to note here that  the spin contribution to the current does not change the Drude resistivity for a given electronic density.)} 
A reasonable approximation is $mv(t) =  -eE\tau(1-e^{-t/\tau})
$, which produces an Edelstein field 
\be
B_E(t)=  -\alpha eE\tau(1-e^{-t/\tau})\,.
\ee
The problem is now to calculate the spin dynamics of the electrons in this time-dependent field, which is no longer linear.  If $\tau$ is sufficiently long the results will be indistinguishable (for $t \ll \tau$)  from those obtained in the previous section.  The non linearity will be observable if  the terminal velocity of the electrons is comparable to the Fermi velocity.  Alternatively, one could have a very large electric field acting on a system with a not-so-large $\tau$.

The inverse Edelstein effect is more delicate.  The charge current will now have contributions not only from the injected spin ($\alpha S_y$), but also from $p_x$. The latter arises because the applied field $B_y$ changes the distribution of the electrons in momentum space.  In fact, under equilibrium conditions the $p_x$ contribution would exactly cancel the $\alpha S_y$ contribution.  One way to  calculate the effect is to solve the spin dynamics in the presence of a linearly growing field $B_y = \Gamma t \Theta(t)$ in the absence of impurities (which gives us the already calculated current $\alpha S_y$)  and then take into account the impurities by subtracting the $p_x$ current generated by the shift in the momentum distribution.  For the latter, in the spirit of the relaxation time approximation, we assume that it is the equilibrium distribution in a  ``retarded" magnetic field $\Gamma[ t - \tau (1-e^{-t/\tau})]$.  This reduces to the clean result in the limit $\tau \to \infty$ (since the external field 
vanishes for negative times and the equilibrium distribution carries then no $p_x$ current).  Whereas in the steady-state regime $t \gg \tau$ it yields a result proportional to $\tau$ as expected from Onsager reciprocity.  Once again, we conclude that the nonlinear effect can be observed if the system is sufficiently clean. 

As a final point we wish to comment on what happens in the case that the electrons are in a Bloch band with periodic dispersion.  The Bloch wave vector $\kv$ is still a constant of the motion.  The Edelstein field oscillates in time at the Bloch frequency $\omega_B=\frac{eEa}{2\pi}$, where $a$ is the lattice constant.  This will induce oscillations in both the Edelstein spin polarization and the spin current.  Depending on whether $\omega_B$ is small or large relative to $\alpha p_F$ we will have adiabatic or non-adiabatic response.  

{\it Acknowledgments} --- This work was supported by NSF grant DMR-1406568 (GV) and by the Donostia International Physics Center (GV) where part of this work was completed.   IVT acknowledges support from the Spanish Grant FIS2013-46159-C3-1-P,  and from the ``Grupos Consolidados UPV/EHU del Gobierno Vasco'' (Gant No. IT578-13)

\appendix*
\section{Solution of Eqs.~(\ref{Equations})}

To see that the functions defined by Eqs.~(\ref{Sol1}) are the solution to Eqs.~(\ref{Equations}) we introduce a rescaled time variable $z=\sqrt{2}e^{-i\frac{3\pi}{4}}\tau$ and rewrite Eqs.~(\ref{Equations}) as follows
\begin{eqnarray}
 \label{Eqs1}
&&\frac{d}{d z} u_\pv(z) + \frac{z}{2}u_\pv(z) + e^{i\frac{\pi}{4}}\frac{\Delta_\pv}{\sqrt{2}}v_\pv(z) = 0 , \nn\\
&& \frac{d}{d z}u_\pv(z) - \frac{z}{2}u_\pv(z) + e^{i\frac{\pi}{4}}\frac{\Delta_\pv}{\sqrt{2}}u_\pv(z) = 0.
\end{eqnarray}
Next, we define a parameter $\nu=-i\Delta_\pv^2/2$ and transform these equations to the form
\begin{eqnarray}
 \label{Eqs2}
&& \frac{d}{d z}u_\pv(z) + \frac{z}{2}u_\pv(z) - \nu \frac{v_\pv(z)}{\sqrt{-\nu}} = 0 , \nn\\
&&\frac{d}{d z}\frac{v_\pv(z)}{\sqrt{-\nu}}  - \frac{z}{2}\frac{v_\pv(z)}{\sqrt{-\nu}} + u_\pv(z)=0.
\end{eqnarray}
By comparing Eqs.~(\ref{Eqs2}) with the following recursion relations for the parabolic cylinder functions $D(\nu,z)$ [see, for example, Refs.~\onlinecite{Bateman-Erdelyi_V2,Gradshtein}]
\begin{eqnarray}
 \label{Recursion}
&&\frac{d}{d z}D(\nu,z)  + \frac{z}{2}D(\nu,z) - \nu D(\nu-1,z)= 0 , \nn\\
&&\frac{d}{d z}D(\nu-1,z)  - \frac{z}{2}D(\nu-1,z) + D(\nu,z)=0,
\end{eqnarray}
we identify the functions $u_\pv(z)=D(\nu,z)$ and $v_\pv(z)=\sqrt{-\nu}D(\nu-1,z)$ as the solutions to Eqs.~(\ref{Eqs1}) and therefore to Eqs.~(\ref{Equations}). By returning to the original time variable $\tau$ we recover Eqs.~(\ref{Sol1}) up to the normalization factor.

To construct the second linear independent solution we notice the following property of Eqs.~(\ref{Equations}). If a pair $\{u_\pv(\tau),v_\pv(\tau)\}$ is a solution to Eqs.~(\ref{Equations}), then $\{-v_\pv^*(\tau),u_\pv^*(\tau)\}$ is also a solution. Moreover these two solution are orthogonal to each other at any $\tau$. Using this property one obtains the solution of Eqs.~(\ref{Sol2}) from Eqs.~(\ref{Sol1}).

\bibliography{SJ2.bib}
\end{document}